\documentclass[12pt]{article}
\usepackage{amsfonts,amsmath,amssymb,subfigure,epsf,psfig}
\usepackage{amsthm,amscd}
\usepackage[dvips]{graphicx}
\makeatletter
\def\@cite#1#2{$^{\hbox{\scriptsize{#1\if@tempswa , #2\fi}}}$}

\@addtoreset{equation}{section}
\makeatother
\theoremstyle{definition}

\newcommand{\zbib}[2]{\frac{d {#1}}{d {#2}}}

\newcommand{\bib}[2]{\frac{\partial {#1}}{\partial {#2}}}
\newcommand{\obib}[2]{\frac{d {#1}}{d {#2}}}

\begin{document}

\title{Yang-Mills instantons on 7-dimensional manifold\\
of $G_2$ holonomy}

\author{
\thanks{E-mail address : miyagi@sci.osaka-cu.ac.jp} 
\vspace*{0.5cm}
Sayuri Miyagi \\
\vspace*{0.1cm}
{\em Department of Physics, Osaka City University},\\
 {\em  Sumiyoshiku, Osaka 558-8585, Japan}}
\date{}
\maketitle
\vspace*{-0.3cm}

\begin{abstract}
We investigate Yang-Mills instantons on 
a 7-dimensional manifold of $G_{2}$ holonomy.
By proposing  a spherically symmetric  
ansatz for the Yang-Mills connection,
we have  ordinary differential equations
as the reduced  instanton equation, 
 and give some explicit and 
numerical solutions.
\quad \\
\quad \\
\end{abstract}
\vspace*{-0.8cm}
\section{Introduction}
\vspace*{-0.5pt}
\noindent
Recent progress in string duality has 
stimulated a great deal of interest in Yang-Mills instantons
on higher dimensional manifolds of special holonomy.\cite{Hull,Kanno0}
The higher dimensional Yang-Mills instanton
equations on such manifolds have been given in
several papers.\cite{Corrigan}${}^{-}$\cite{Kanno}
It is known that some Yang-Mills instantons on
the manifolds can be extended to  solitonic solutions of the 
 low energy effective theory
of the heterotic string\cite{Acharya3,GN}
preserving $ \frac1{16}$ or $ \frac2{16}$
of the 10-dimensional $N=1$ supersymmetry. 

The manifolds of special holonomy admit covariantly constant spinors 
and define the supersymmetric cycles
 (calibrated submanifolds).\cite{Kanno}
Solitons and instantons of superstring theory or M-theory
can be considered to be wrapping branes around a 
supersymmetric cycle.\cite{BBS,Ooguri,Becker} 
The manifolds also play an important role in
generalizations of the Donaldson-Floer-Witten theory
to higher dimensions.\cite{BKS,Acharya2}
Baulieu et al\cite{BKS} found a topological action
on a 8-dimensional manifold
of $Spin(7)$ holonomy; their covariant gauge fixing 
conditions lead to Yang-Mills instanton equations on the manifold. 
Compared with the 4-dimensional case,
there are a few explicit Yang-Mills instantons on
7- and 8-dimensional manifolds of $G_2$
and $Spin(7)$ holonomy.\cite{Acharya3,GN,FN}
Recently  Yang-Mills instantons on a 8-dimensional manifold
of $Spin(7)$ holonomy have been examined in detail.\cite{KY}

In this paper  we investigate Yang-Mills instantons
on a 7-dimensional manifold of $G_2$ holonomy.
Our paper is organized as follows: 
in section 2 and section 3 we briefly review the octonionic algebra
 and the explicit metric of a manifold of $G_2$
holonomy given by Gibbons et al;\cite{GPP} in section 4 
we express Yang-Mills instantons on 7-dimensional manifolds 
of $G_2$ holonomy; in section 5, under a spherically 
symmetric ansatz, we solve the Yang-Mills instanton equations 
on the manifold described in section 3, and present
some new  explicit solutions obtained from the Riccati equation
and numerical solutions.


\section{Octonionic Algebra}
\noindent
In this section we  briefly review the octonionic algebra
following the papers.\cite{GN,FN,koushiki} 
The exceptional Lie group $G_{2}$ which is a subgroup of $Spin(7)$ 
is the automorphism group of the octonionic algebra.
The generators of $Spin(7)$ are given by 
\begin{equation}
    \Gamma^{ab} =\frac12 ( \Gamma^{a} \Gamma^{b} - \Gamma^{b} \Gamma^{a}),
\end{equation}
where $ \Gamma^a(a=1,\cdots,7)$ are 7-dimensional $\gamma$ matrices defined by
\begin{equation}
\{ \Gamma^a, \Gamma^b \} = 2 \delta^{ab}.
\end{equation}
The generators of the octonionic algebra 
$\{{\bf 1}, e_a \},(a=1,\cdots , 7)$ obey the relations
\begin{equation}
e_a \cdot e_b =  - \delta_{ab} {\bf 1} + C_{abc} e_c,
\end{equation}
where $C_{abc}$ is totally antisymmetric tensor with non zero components
\begin{equation}
C_{123}= C_{516} = C_{624} = C_{435} = C_{471} = C_{673} = C_{572} =+1.
\end{equation}
If the  dual tensor $C_{abcd}$ is defined by
\begin{equation}
C_{abcd} = \frac{1}{3!} {\epsilon_{abcd}}^{ijk} C_{ijk}~,
\end{equation}
the following relations are satisfied
\begin{eqnarray}\quad
C^{abpq} C_{cdpq}& =& 4 (\delta^a_c\delta^b_d
- \delta^a_d\delta^b_c) -2 {C^{ab}}_{cd},\label{relation}\\
C^{apq}C_{bcpq} &=& -4 {C^{a}}_{bc}.\label{relation2}
\end{eqnarray}
By the use of the tensor $C_{abcd}$, 
one can decompose the 21-dimensional Lie algebra of $Spin(7)$
into the direct sum of the 14-dimensional Lie algebra of $G_2$ and the
7-dimensional orthogonal subspace ${\cal P}$.  
Indeed, the projector ${{P_1}^{ab}}_{cd}$ onto the Lie algebra
of $G_2$ and  the projector ${{P_2}^{ab}}_{cd}$ onto ${\cal P}$
are given by
\begin{eqnarray}
  \label{eq:projection}
 && {{P_1}^{ab}}_{cd} = \frac23 \left( \frac12 (\delta^a_c \delta^b_d
-\delta^a_d \delta^b_c) + \frac14 {C^{ab}}_{cd}\right),\\
 && {{P_2}^{ab}}_{cd} = \frac13 \left( \frac12 (\delta^a_c \delta^b_d
-\delta^a_d \delta^b_c) - \frac12 {C^{ab}}_{cd}\right).
\end{eqnarray}
Thus the generators $G^{mn}$ of $G_{2}$ can be realized as\cite{GN} 
\begin{equation}
  G^{ab}= \frac34{{P_1}^{ab}}_{cd} \Gamma^{cd}=
\frac12 (\Gamma^{ab} + \frac14 {C^{ab}}_{cd} \Gamma^{cd}).
\end{equation}

\section{Explicit metric on 7-dimensional manifold of $G_{2}$
  holonomy}\label{GI}
\noindent
In this section we describe the explicit metric on a 7-dimensional
manifold of $G_{2}$ holonomy.
It has been  first given as a metric on the $R^4$ bundle
over $S^3$ by solving the Einstein equation:\cite{GPP}
\begin{equation}\label{Ginst}
 ds^2 =
 \alpha^2 dr^2 + \beta^2 (\sigma^i - A^i )^2 +
\gamma^2 \Sigma^i\Sigma^i,
\end{equation}
where $i=1,2,3\,$
 and  $\alpha, \beta, \gamma$ are functions solely of the radial
coordinate $r$. The symbols $\Sigma^i$ denote the left invariant 1-forms 
on the $S^3$ base manifold and the symbols $\sigma^i$ denote the left invariant
1-forms on the fiber $R^4$. They satisfy
\begin{equation}
d \Sigma^i = -\frac12 \epsilon_{ijk} \Sigma^j \wedge \Sigma^k,\quad
d \sigma^i = -\frac12 \epsilon_{ijk} \sigma^j \wedge \sigma^k.  
\end{equation}
Then $SU(2)$ connection $A^i$ is given by $A^i = \frac12
\Sigma^i$ and 
the explicit forms of the functions are
\begin{equation}\label{kai}
  \alpha^2=\left(1-\left({\frac{a}{r}}\right)^{3}\right)^{-1} , \quad
\beta^2= \frac19 r^2 \left(1-\left({\frac{a}{r}}
  \right)^{3}\right)
, \quad \gamma^2 =\frac{\displaystyle r^2}{12}
\end{equation}
where $a$ is an arbitrary constant.

We can also construct the metric (\ref{Ginst})  
by solving the first order equation for
the spin connection,\cite{RF,Bakas}
\begin{equation}
 \omega_{ab}= \frac12 {C_{ab}}^{cd} \omega_{cd}\label{spininst}.
\end{equation}
Under the spherically symmetric ansatz (\ref{Ginst}), the orthonormal basis is 
\begin{equation}
  e^7 = \alpha d r , \quad e^i = \gamma \Sigma^i , \quad 
e^{\hat{i}} = \beta(\sigma^i -A^i)
\end{equation}
where $\hat{i}= \hat{1},\hat{2},\hat{3}=4,5,6$, 
and the spin connection $\omega_{ab}$ is obtained by the equations 
$d e^a ={ \omega^a}_{b} \wedge e^b\label{spin}$ and $\omega_{ab} = - \omega_{ba}$: 
\begin{eqnarray}
&&\omega_{7i}= - \frac{\gamma^{'}}{\alpha\gamma} e^i,\label{spinconnb}\\
&&\omega_{7\hat{i}}= - \frac{\beta^{'}}{\alpha\beta} e^{\hat{i}},\\
&&\omega_{i\hat{j}}=  \frac{\beta}{8\gamma^2}\epsilon_{ijk} e^{k},\\
&&\omega_{ij} = - \epsilon_{ijk}(\frac1{2 \gamma}e^k 
+ \frac{\beta}{8 \gamma^2} e^{\hat{k}}),\\
&&\omega_{\hat{i}\hat{j}} = - \epsilon_{ijk}(\frac1{2 \gamma}e^k 
+ \frac{1}{2 \beta} e^{\hat{k}}).\label{spinconne}
\end{eqnarray}
Eq. (\ref{spininst}) then reduces to  the 
following nonlinear ordinary equations:
  \begin{equation}
    \label{eq:abc}
\beta^{'}=\frac{\alpha \beta^2}{8 \gamma^2} -\frac{\alpha}{2}
 , \quad \gamma^{'}= - \frac{\alpha \beta}{4 \gamma},  
  \end{equation}
which reproduce the solution (\ref{kai}).

\section{Yang-Mills instanton on 7-dimensional manifold of $G_2$ holonomy}
\noindent
Here we define a Yang-Mills instanton equation on the
7-dimensional manifold of $G_2$ holonomy in terms of 
a special closed $3$-form $\Omega$,\cite{Corrigan,Kanno}
\begin{equation}
  \Omega \wedge F =  * F.\label{inst}
\end{equation}
Using the antisymmetric tensor $C_{abc}$, 
the 3-form $\Omega$ can be written as
\begin{equation}
  \label{eq:omega}
  \Omega= \frac{1}{3!} C_{abc} e^a \wedge e^b \wedge e^c,
\end{equation}
and then Eq. (\ref{inst}) becomes 
\begin{equation}
  \label{YMinst}
  F_{ab}= \frac12 {C_{ab}}^{cd} F_{cd}.
\end{equation}
The seven components of  Eq. (\ref{YMinst}) are as follows:  
\begin{eqnarray}
&&  F_{71}=  F_{26}+ F_{53},\,
    F_{72}=  F_{61}+ F_{34},\,
    F_{73}=  F_{42}+ F_{15},\nonumber\\
&&  F_{74}=  F_{23}+ F_{65},\,
    F_{75}=  F_{46}+ F_{31},\,
    F_{76}=  F_{12}+ F_{54},\\
&&  F_{63}=  F_{25}+ F_{14}.\nonumber
\end{eqnarray}
The curvatures (\ref{YMinst}) provide 
the solutions of the Yang-Mills equation on the
manifold, $D_{a} F^{ab}=0$,  as a consequence of the Bianchi identity. 
We shall call solutions of (\ref{YMinst})
Yang-Mills instantons.
The Yang-Mills instantons can be embedded into supersymmetric theories
as solutions to the equation of motion.
It is known that there is a spinor $\eta$
on the manifold of $G_2$ holonomy
which satisfies the conditions:
\begin{equation}
\label{chii}
  \nabla \eta =0 ,  \quad  G^{ab} \eta = 0,
\end{equation}
where $\nabla$ denotes 
the covariant derivative of the Levi-Civita connection
on the manifold.\cite{Corrigan,Bakas}
Then, one can use the covariantly constant spinor as a global supersymmetric 
parameter defined on the manifold.
The supersymmetric transformation of a spinor field $\chi$,
a super partner of Yang-Mills connection,
\begin{equation}
  \label{SYM}
  \delta \chi = \frac12 F_{ab} \Gamma^{ab} \eta 
= \frac12 ({{P_{1}}^{ab}}_{cd}+{ {P_{2}}^{ab}}_{cd}) F_{ab}
\Gamma^{cd} \eta
= \frac23 F_{ab} G^{ab} \eta + \frac16 ( F_{ab}- \frac12 {C_{ab}}^{cd}
F_{cd}) \Gamma^{ab} \eta 
\end{equation}
is zero according to the Eqs. (\ref{YMinst}) and (\ref{chii}).
So the Yang-Mills instanton becomes a supersymmetric solution  
preserving  the  spinor $\eta$.

Note that the spin connection leads to 
a $G_2$ Yang-Mills connection:
\begin{equation}\label{sa}
  {\cal A} = \frac13 G^{ab}{\omega_{ab}}.
\end{equation}
Using (\ref{spininst}) and the symmetry between the first  
and second pair of indices of the Riemann curvature $R_{abcd}$,
 we can see that the curvature of ${\cal A}$ satisfies the Yang-Mills instanton equation 
(\ref{YMinst}).\cite{Acharya3}

\newpage
\section{Ansatz for $G_2$ Yang-Mills connection}
\noindent
Now we construct Yang-Mills
instantons on the 7-dimensional manifold 
of $G_2$ holonomy described in section 3.
Referring to the spin connection (\ref{spinconnb})-(\ref{spinconne}),
we propose a spherically symmetric  ansatz for the $G_2$ Yang-Mills
connection $ {\cal A}= \frac13 G^{ab} A_{ab}$:
\begin{eqnarray}
&&A_{7i}=2A e^i\label{Ab},\\
&&A_{7\hat{i}}=(B-C)e^{\hat{i}},\\
&&A_{i\hat{j}}=\epsilon_{ijk}A e^k,\\
&&A_{ij}=\epsilon_{ijk}(D e^k+ B e^{\hat{k}}),\\
&&A_{\hat{i}\hat{j}}=\epsilon_{ijk}(D e^k+ C e^{\hat{k}}),\label{Ae}
\end{eqnarray}
where $A, B, C, D$ are the functions of the radial coordinate  $r$. 
We can see that the above connection satisfies
\begin{equation}\label{Finst}
A_{ij} =\frac12 {C_{ij}}^{kl} {A_{kl}}
\end{equation}
and so does their curvature, i.e., they are just projected onto 
the Lie algebra of $G_2$. Note that the indices denote 
those of the Lie algebra of $G_2$, not of the differential forms.
This connection reproduces Eq. (\ref{sa}) by putting
\begin{equation}
A=\frac{\beta}{8 \gamma^2}, \quad B= - \frac{\beta}{8 \gamma^2}, \quad
C=-\frac{1}{2\beta}, \quad D=-\frac1{2 \gamma}.  
\end{equation}

The components of the curvature  ${\cal F} (= \frac13G^{ab}F_{ab})$ 
 are explicitly given by
\begin{eqnarray}
  \label{eq:curvature}
F_{7i}&=& \frac2{\alpha}\! (A^{'} \!+\! \frac{\gamma^{'}}{\gamma} A)
e^7 \!\wedge\! e^i \! -\epsilon_{ijk} A (\frac1{\gamma} + 2 D )
 e^j\! \wedge\! e^k\! -\!\epsilon_{ijk}\! A  (3B -C ) e^j \wedge e^{\hat{k}} ,\\
F_{7 \hat{i}}&=& \frac1{\alpha} \bigl( (B-C)^{'} +
  \frac{\beta^{'}}{\beta} (B-C) ) e^7 \wedge e^{\hat{i}}-
\epsilon_{ijk} (B-C) (\frac1{2\beta} + C\bigr) e^{\hat{j}} \wedge e^{\hat{k}}\nonumber\\
&&\!\!\!-\epsilon_{ijk} (B-C) (\frac1{2\gamma} + D) 
e^{\hat{j}} \wedge e^{k}
+ \epsilon_{ijk} \bigl( \frac{\beta}{8\gamma^2} ( B-C) -2 A^2 \bigr)
e^j \wedge e^k, \! \! \\
 F_{i \hat{j}}& =& \epsilon_{ijk} \frac1{\alpha} (A^{'} + \frac{\gamma^{'}}{\gamma} A)
e^7 \wedge e^k -A (\frac1{\gamma} + 2 D ) e^i \wedge e^j
-A ( 3B-2C ) e^i \wedge e^{\hat{j}}\nonumber\\
&& - AC e^{\hat{i}} \wedge e^j 
+ \delta_{ij} A (B-C)  e^l   \wedge e^{\hat{l}},\\
F_{ij} &=& \epsilon_{ijk} \frac1{\alpha} (D^{'} + \frac{\gamma^{'}}{\gamma} D)
e^7 \wedge e^k 
+\epsilon_{ijk} \frac1{\alpha} (B^{'} + \frac{\beta^{'}}{\beta} B)
e^7 \wedge e^{\hat{k}}\nonumber\\
&&-( \frac{D}{\gamma} - \frac{\beta}{4 \gamma^2} B +5 A^2 + D^2 ) 
e^i \wedge e^j  -B( \frac1{\beta} + B) e^{\hat{i}} 
\wedge e^{\hat{j}}\nonumber \\
&&- B ( \frac1{2 \gamma} + D) e^{\hat{i}} \wedge e^j
+ B ( \frac1{2 \gamma} + D) e^{\hat{j}} \wedge e^i,
\end{eqnarray}
\begin{eqnarray}
F_{\hat{i} \hat{j}}& =& 
\epsilon_{ijk} \frac1{\alpha} (D^{'} + \frac{\gamma^{'}}{\gamma} D)
e^7 \wedge e^k 
+\epsilon_{ijk} \frac1{\alpha} (C^{'} + \frac{\beta^{'}}{\beta} C)
e^7 \wedge e^{\hat{k}}\nonumber\\
&&-( \frac{D}{\gamma} - \frac{\beta}{4 \gamma^2} C +A^2 + D^2 ) 
e^i \wedge e^j -( \frac{C}{\beta} + (B-C)^2 + C^2 ) e^{\hat{i}} 
\wedge e^{\hat{j}}\nonumber\\
&&- C ( \frac1{2 \gamma} + D) e^{\hat{i}} \wedge e^j
+ C ( \frac1{2 \gamma} + D) e^{\hat{j}} \wedge e^i.
\end{eqnarray}

Let us consider the Yang-Mills instanton equation (\ref{YMinst})
for the curvature $F_{ab} = \frac12 F_{abij} e^i \wedge e^j$:
\begin{equation}\label{Finst}
F_{abij} =\frac12 {C_{ij}}^{kl} {F_{abkl}}.
\end{equation}
After some calculations, we obtain a set of 
ordinary differential equations for $A,B,C,D$: 
\begin{eqnarray}
&&(B-C)\bigl(\frac{1}{2 \gamma} +D \bigr)=0, 
\label{eq:ABCD2}\\
&&D^{'} + \frac{\gamma^{'}}{\gamma} D = -2 B \alpha(\frac1{2\gamma}+D), 
\label{eq:ABCD3} \quad
D^{'} + \frac{\gamma^{'}}{\gamma} D = -2 C \alpha\bigl(\frac1{2\gamma}+D\bigr),
 \label{eq:ABCD4}\\
&&  A^{'}+\frac{\gamma^{'}}{\gamma} A = - A \alpha(3B - C),
\quad A(\frac1{\gamma}+ 2D)=0,
\label{eq:ABCD1}\\
&&(B-C)^{'} + \frac{\beta^{'}}{\beta} (B-C) = 
\alpha \bigl(\frac{\beta}{4 \gamma^2} (B-C)-4 A^2 \bigr)
+ 2 \alpha (B-C)\bigl(\frac1{2 \beta} + C\bigr)
\label{eq:ABCD6},\\
&&-\frac1{\alpha}\bigl(B^{'}+ \frac{\beta^{'}}{\beta} B\bigr)=
\frac{D}{\gamma} - \frac{\beta B}{4 \gamma^2} + 5 A^2
 +D^2 - \frac{B}{\beta}-B^2, \label{eq:ABCD8}\\
&&-\frac1{\alpha}\bigl(C^{'}+ \frac{\beta^{'}}{\beta} C\bigr)=
\frac{D}{\gamma} - \frac{\beta C}{4 \gamma^2} + A^2
 +D^2 - \bigl(\frac{C}{\beta}+ (B-C)^2 + C^2\bigr). \label{eq:ABCD9}
\end{eqnarray}
Eq. (\ref{eq:ABCD2}) can be satisfied in
three cases (i) $ D = -\frac1{2 \gamma}, B \ne C$,  
(ii) $D=-\frac1{2 \gamma}, B=C, $
(iii)$  D  \ne  -\frac1{2 \gamma}, B=C,$
 so we study these cases separately.

In case (i), we  have two solutions:
\begin{equation}
(A,B,C,D)=\left(\frac{\beta}{8\gamma^2},
-\frac{\beta}{ 8 \gamma^2 },-\frac{1}{2 \beta},-\frac{1}{2\gamma}\right), 
\quad     
\left(-\frac{\beta}{8 \gamma^2},
-\frac{\beta}{8 \gamma^2},-\frac{1}{2 \beta},-\frac{1}{2\gamma}\right).
\end{equation}
The first one is nothing but the connection (\ref{sa}), while
the second one is a new solution.
Although their difference is only the signature of $A$,
these two solutions lead to different curvatures. 

In cases (ii) and (iii), $A$ is zero from (\ref{eq:ABCD6}) 
so that the Yang-Mills instanton equation (\ref{YMinst})
gives two nonlinear ordinary differential equations for $B$ and $D$:
\begin{eqnarray}
&& B^{'}= \alpha \left( - \frac{1+ 9 \alpha^2}{2 \alpha r}B
- \frac{2 \sqrt{3}}{r} D - D^2 + B^2 \right)\label{eq:reduction1},\\
&& D^{'} = 
\alpha \left( - \frac{D}{\alpha r} - \frac{2\sqrt{3}}{r} B 
-2 B D \right)\label{eq:reduction},
\end{eqnarray}
where we have used Eqs. (\ref{eq:abc}) and (\ref{kai}).
Let us introduce  new variables $X$ and $Y$,
\begin{equation}\label{ct}
  B=\frac{X+1}{r},\quad  D=\frac{Y-\sqrt{3}}{r}
\end{equation}
and a new parameter $s= \mbox{ln}(\frac{r}{a})$. 
Then Eqs. (\ref{eq:reduction1}) and (\ref{eq:reduction}) become
\begin{eqnarray}
 && \zbib{X}{s} = \frac{1-9 \alpha^2}{2}(X+1)
+3\alpha - \alpha  Y^2 +\alpha(X+1)^2,
\label{eq:mujigen1}\\
 && \zbib{Y}{s}= -2\alpha (X+1) Y \label{eq:mujigen2}.
\end{eqnarray}

In case  (ii), the Eq. (\ref{eq:mujigen2}) is trivial as
 $Y=0 \, ( D = -\frac1{2 \gamma})$ 
and  Eq. (\ref{eq:mujigen1}) becomes the Riccati equation:
\begin{equation}
  \label{eq:X}
  \zbib{X}{s} = \frac{1-9 \alpha^2}{2}(X+1) +3 \alpha
 +\alpha (X+1)^2.
\end{equation}
We can find a general solution of  Eq. (\ref{eq:X}):
\begin{equation}\label{Xsol}
X=3\alpha -1 + \frac{1}{\alpha(-\frac12 + \frac{c}{r^2} )},
\end{equation}
where $c$ is an arbitrary integration constant. 
Taking the limit  $ c \to \infty $, we have a special solution
$X= 3\alpha-1$. For the explicit solution (\ref{Xsol}) 
the nonzero components of the Yang-Mills connection and
 the curvature are $\{A_{ij}, A_{\hat{i} \hat{j}}=A_{ij}\}$
and $\{F_{ij}, F_{\hat{i} \hat{j}}=F_{ij}\}$.
In this case the curvature becomes
\begin{equation}
 {\cal F} = \frac13 G^{ab} F_{ab} = \frac13 (G^{ij} +  G^{\hat{i} \hat{j}})F_{ij}
= \frac13 T^{ij} F_{ij},
\end{equation}
where $T^{ij}$ are $SU(2)$ generators; 
the gauge group reduces to $SU(2)$ from $G_2$.
The explicit components of the curvature become
\begin{eqnarray}
&&F_{ij7 \hat{k}}= F_{\hat{i}\hat{j}7\hat{k}}=
\epsilon_{ijk}\frac1{r^2}
\left(\frac{2 + (\frac{a}{r})^3}{-\frac12 + \frac{c}{r^2}}
+ \frac{1}{(-\frac12  + \frac{c}{r^2})^2 \alpha^2}\right),\label{F1}\\
&&F_{ijij} = F_{\hat{i}\hat{j}ij}=
-\frac1{r^2} \frac{1}{(-\frac12  + \frac{c}{r^2})
  \alpha^2},\\
&&F_{ij\hat{i}\hat{j}} = F_{\hat{i}\hat{j}\hat{i}\hat{j}}
=-\frac1{r^2}
 \left(\frac{3}{-\frac12  + \frac{c}{r^2}}
+ \frac{1}{(-\frac12 + \frac{c}{r^2})^2 \alpha^2} \right).\label{F3}
\end{eqnarray}

It is known that there are no Yang-Mills instantons
with finite action on the n-dimensional flat 
 space (or sphere) unless $n = 4.$\cite{Kanno}
Unfortunately, the action  for our solution diverges too. 
However the second Chern class 
evaluated on the fiber $R^{4}$ 
 has a finite value:
\begin{eqnarray}
 c_2& =&\int_{R^4} Tr \,{\cal F} \wedge {\cal F} \nonumber\\
&=& -\frac49  \, \, \mbox{vol}_{SU(2)} \int_{a}^{\infty} \frac{d}{dr} \left(
\frac1{3\alpha^6} (-\frac12 + \frac{ c}{r^2})^{-3} +
 \frac3{2\alpha^4} (-\frac12 + \frac{c}{r^2})^{-2}\right) d r  \nonumber\\
&=&-\frac{40}{27} \,  \mbox{vol}_{SU(2)},
\end{eqnarray}
where $Tr$ refers to the  adjoint representation of 
$SU(2)$ 
and  $\mbox{vol}_{SU(2)}$ is the volume of  $SU(2)$. 
Note that the value comes from the infinite surface
according to $\frac{1}{\alpha}=0 $ at $r=a$.
The meaning of the finite second Chern class remains
an open problem here.

Now we proceed to case (iii).
We first note that the two points on the  $(X,Y)$-plane, 
$L=(-1, \sqrt{3} ), M=(-1, - \sqrt{3})$, 
are fixed points, i.e., the stationary solutions of 
 Eqs. (\ref{eq:mujigen1}) and (\ref{eq:mujigen2}). 
We calculate other solutions numerically
 and show the flows of appropriate numerical solutions 
in Figure 1.

\begin{figure}[hhhh]
  \begin{center}
\leavevmode
\epsfxsize=7cm  
\epsffile{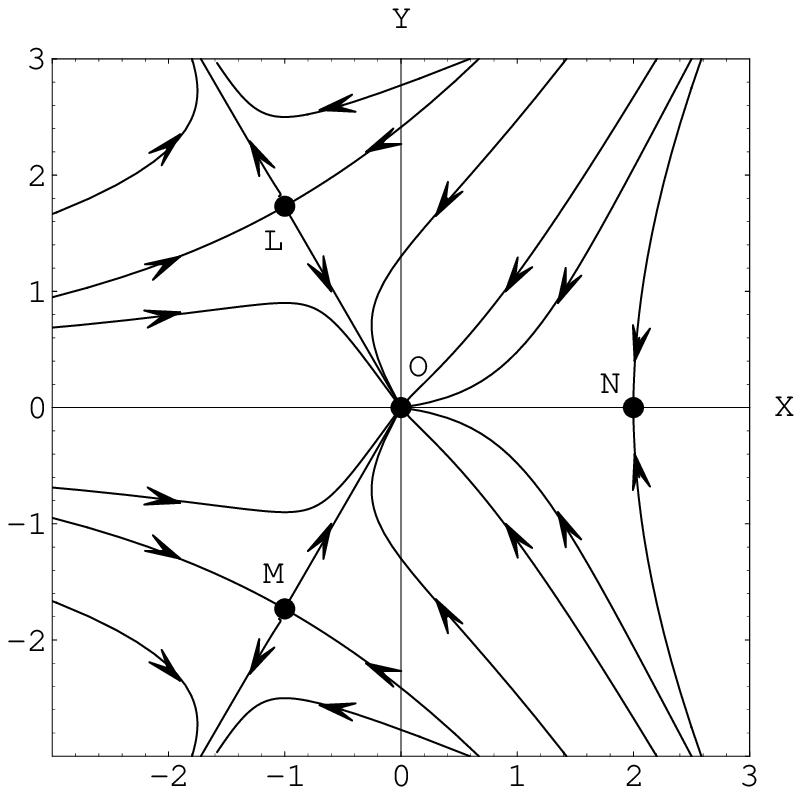} 
  \caption{Flows of the numerical solutions in the case of (iii).}
{\quad The  arrows show the growing direction of $r$.}
  \end{center}
\end{figure}
We can see  general features of the solutions 
if we rewrite Eqs. (\ref{eq:mujigen1}) and (\ref{eq:mujigen2}) as 
\begin{eqnarray}
&&  \obib{{ X}}{{ s}} 
= -\bib{{ W}_{\alpha}}{{ X}},\quad \quad  
\obib{Y}{s} = -\bib{W_{\alpha}}{Y},\label{bibXY}\\
  && W_{\alpha}= -\left(
\frac{1-9 \alpha^2}{4}(X+1)^2 + \alpha (3-Y^2)X + \frac{\alpha}{3} 
(X+1)^3  - \alpha Y^2 \right)\label{W}.
\end{eqnarray}
Note that the external field $\alpha$ coming from the background metric
 is included in the equations. 
If we consider the asymptotic region (large radial coordinate region), 
we can regard  Eqs. (\ref{bibXY}) and (\ref{W})
as gradient flow equations since  $\alpha$ approaches a constant.
Then two fixed points,  
$O=( 0 , 0 ), N=( 2 ,0)$,  appear, adding to the points $ L$ 
and $M$ described above.  
By evaluating the $2 \times 2 $   Hessian matrix $H$
\begin{equation}\label{H}
  H_{ij}= \frac{\partial^2 W_1}{\partial X^i \partial X^j} 
\quad (X^1= X, X^2=Y) 
\end{equation}
at the four fixed points, we can see  that 
$O$ is a stable point and $L,M,N$ are saddle points 
(the Morse index is $1$). 
This analysis  
provides a qualitative explanation for the behavior of
the flows in Figure 1.

\section*{\bf Acknowledgments}
\noindent
I would like to thank Y. Yasui, Y. Hashimoto, H. Kanno,
T. Ootsuka and T. Takaai for helpful discussions.



\begin{thebibliography}{99}



\bibitem{Hull}C. M. Hull,
{\em Adv. Theor. Math. Phys.} {\bf 2}, 619 (1998)

\bibitem{Kanno0}L. Baulieu, H. Kanno and I. M. Singer,
hep-th/9705127



\bibitem{Corrigan}E. Corrigan, C. Devchand and D. B. Fairlie,
{\em Nucl. Phys.} {\bf B214}, 452 (1983).

\bibitem{Ward}R. S. Ward,
{\em Nucl. Phys.} {\bf B236}, 381 (1984).

\bibitem{Popov}A. D. Popov,
{\em Mod. Phys. Lett.} {\bf A23} 2077 (1992).

\bibitem{Kanno}H. Kanno,
hep-th/9903260.


\bibitem{Acharya3}B. S. Acharya and M. O'Loughlin,
{\em Phys. Rev.} {\bf D55}, 4521 (1997).



\bibitem{GN}M. G\"unaydin and H. Nicolai,
{\em Phys. Lett.} {\bf B351}, 169 (1995).




\bibitem{BBS}K. Becker, M. Becker and A. Strominger,
{\em Nucl. Phys.} {\bf B456}, 130 (1995).

\bibitem{Ooguri}H. Ooguri, Y. Oz and Z. Yin,
{\em Nucl. Phys.} {\bf B477}, 407 (1996).

\bibitem{Becker}K. Becker, M. Becker, D. R. Morrison, H. Ooguri, Y. Oz
and Z. Yin,
 {\em Nucl. Phys.}  {\bf B480}, 225 (1996).

\bibitem{BKS}L. Baulieu, H. Kanno and I. M. Singer,
{\em Comm. Math. Phys.} {\bf 194}, 149 (1998)

\bibitem{Acharya2}B. S. Acharya and M. O'Loughlin and B. Spence,
{\em Nucl. Phys.} {\bf B503}, 657 (1997).


\bibitem{FN}S. Fubini and H. Nicolai,
{\em Phys. Lett.} {\bf B155}, 369 (1985).


 
\bibitem{KY}H. Kanno and Y. Yasui,
hep-th/9910003.


\bibitem{GPP}G. W. Gibbons, D. N. Page and C. N. Pope, 
{\em Commun. Math. Phys.} {\bf 127}, 529 (1990).


\bibitem{koushiki}R. D\"undarer, F. G\"ursey and C. H. Tze, 
{\em J. Math. Phys.} {\bf 25}, 1496 (1984).


\bibitem{RF}D. Brecher and M. J. Perry,
hep-th/9908018.

\bibitem{Bakas}I. Bakas, E. G. Foratos and A. Kehagias,
{\em Phys. Lett.} {\bf B445}, 9 (1998)





\end{thebibliography}
\end{document}